\documentclass[doublecol]{epl2} 
\usepackage{amsmath}
\usepackage{amssymb}

\title{Renormalisation group determination of the order of the DNA denaturation transition}
\shorttitle{RG determination of the order of the denaturation transition of DNA} %Insert here a short version of the title if it exceeds 70 characters

\author{Jos\'e Manuel Romero-Enrique\inst{1} \and Francisco de los Santos\inst{2,3} \and Miguel A. Mu\~noz\inst{2,3}}
\shortauthor{J.M. Romero-Enrique, F. de los Santos, and M.A. Mu\~noz}

\institute{                    
  \inst{1} Departamento de F\'isica At\'omica, Molecular y Nuclear - Universidad de Sevilla, Apartado de correos 1065, 41080 Seville, Spain\\
  \inst{2} Departamento de Electromagnetismo y F\'isica de la Materia - Universidad de Granada, Fuentenueva s/n, 18071 Granada, Spain\\
  \inst{3} Instituto Carlos I de F\'isica Te\'orica y Computacional - Universidad de Granada, Fuentenueva s/n, 18071 Granada, Spain
}

\abstract{
We report on the nature of the thermal denaturation transition of homogeneous DNA as
determined from a renormalisation group analysis of the Peyrard-Bishop-Dauxois model.
Our approach is based on an analogy with the phenomenon of critical wetting that goes 
further than previous qualitative comparisons, and shows that the transition is 
continuous for the average base-pair separation. However, since the range of universal critical behaviour
appears to be very narrow, numerically observed denaturation transitions
may look first-order, as it has been reported in the literature.
}

\begin{document}
\newcommand\h{\langle h \rangle}
\maketitle

\section{Introduction}
In addition to its central relevance in biology, the DNA molecule also displays a variety of remarkable
physical properties, some of which are key to the understanding of DNA function~\cite{yakushevich}.
For instance, mechanical properties such as bending, twisting, or compression are directly related to 
DNA replication or transcription, which requires that the two strands are separated in order that they
can be read by DNA or RNA polymerase. This can be achieved by various mechanisms, including pulling enzymes, 
mechanical force or gentle heating. 
In this latter case, the process is known as 
DNA thermal denaturation or DNA melting, and has received a great deal of attention over several decades~\cite{zm}.
Experimental observation of the fraction of bound pairs or the average base-pair separation as a function of temperature
reveals a sharp jump in the denaturation curves from double-- to single--stranded DNA, a behaviour that hints at some sort of phase transition.
However, controversies remain regarding the nature of this transition (whether first or second order), see below. 
This point goes beyond mere physical curiosity and has biological relevance, for
there is increasing interest in the correspondence between functional and thermodynamic melting properties, e.g., the identification of coding sequences in genomes on the basis of thermodynamic melting behaviour~\cite{jost}.

Models of varying complexity and applicability have been developed that account for DNA melting.
Two large families are based on the models of Poland-Scheraga (PS)~\cite{PS} and Peyrard-Bishop-Dauxois (PBD)~\cite{PB}, 
or modifications thereof. Within the PS framework, the DNA is described as a sequence of base pairs that can be either bound
or unbound. Thermal fluctuations cause segments of DNA to unbind, creating temporarily denaturated loops of variable size
which can ultimately coalesce upon increasing the temperature, thus 
triggering the denaturation transition. It has been shown that the entropy contribution of the loops depends on their size, $l$, 
as $\sim 1/l^c$, and three different scenarios have been reported depending only on the value of $c$: $c\leq 1$, no phase transition,
$1 < c \leq 2$, continuous transition, and $c>2$, first-order phase transition. The value of $c$ is not easily determined, but
it has been recently demonstrated that taking into account the excluded-volume interactions between denaturated loops and the rest
of the chain is enough to give $c>2$~\cite{kafri}, so the transition is therefore first-order (see \cite{guttmann} for a refined analysis of these
issues and a discussion of some open questions).

Turning to PBD-type models, the situation is much less clear. The PBD model considers only the stretching between corresponding 
base-pairs (see details next section). The transition proceeds as described for the PS, but including intermediate states
because the stretching is a continuously varying variable. It has been thoroughly studied by means of Monte Carlo simulations, Langevin dynamics, 
path integral methods and different transfer integral approaches, but the question of the order of the transition remains as yet unsettled.
Claims have been reported in the literature that the transition is first-order yet with a diverging correlation length,
asymptotically second-order although very sharp looking in appearance, while other studies are inconclusive. 
We shall return to this point in the next section.

In this Letter we take on the question of the order of the DNA denaturation transition in the PBD model
by means of an exact renormalisation group analysis. Our approach is based on an analogy with the 
phenomenon of critical wetting that goes further than previous qualitative comparisons. In the next section,
we review the PB and PBD models. Next, we explain the analogy with wetting and then proceed with 
the renormalisation group calculation, which shows that the transition is continuous for the average
base-pair separation. A summary with our conclusions is presented in the final section.

%See fig.~\ref{fig.1}, table~\ref{tab.1} and eq.~(\ref{eq.1}).
%See also~\cite{b.a,b.b}.

\section{The Peyrard-Bishop-Dauxois model of DNA}
\label{PBD}

The Peyrard-Bishop model ignores the helicoidal structure of the DNA molecule and the properties associated with it, 
and focuses on the stretching of the hydrogen bonds connecting base pairs, which are represented by continuous variables 
$h_n$ ($n=1,2,\ldots, N$, where $N$ is the length of the chain). 
For homogeneous samples (only AT of GC pairs), the Hamiltonian of the model reads~\cite{PB}

\begin{equation}
H=\sum_{n=1}^N \left[
\frac{1}{2}m  p_n^2+
W(h_n,h_{n-1})+
%\frac{k}{2}(h_n-h_{n-1})^2+
V(h_n)
%\frac{1}{2} D (e^{- a h_n}-1)^2
\right],
\label{pb_hamiltonian}
\end{equation}
where the first term is the kinetic energy for bases of mass $m$, $W(h_n,h_{n-1})=k(h_n-h_{n-1})^2$ describes 
the harmonic stacking interaction between neighbouring bases, and $V$ represents the average potential 
between the two bases in a pair which is modeled by a Morse potential, $V(h_n)=D(e^{-ah_n}-1)^2$.
Note that the asymmetry of the two strands is neglected in that a common mass $m$ for the bases is used
and the same stacking coefficient $k$ along the chain is assumed. $D$ is the dissociation energy of the
pairs and $a$ denotes the spatial range of the potential. Their precise values, which are unimportant for
our purposes, can be determined from the fitting of DNA experimental denaturation curves~\cite{Campa}.
Two standard observables are the average stretching $\langle h \rangle$ and the density of bound base pairs $\langle e^{-h} \rangle$.

It was soon recognised that the simple PB model needed to be improved if it was to properly account for the sharp 
shape of the experimental denaturation curves. This was achieved by changing from a harmonic stacking interaction to 
the nonharmonic one
\begin{equation}
%\frac{k}{2} \to \frac{k}{2} \left[ 1+ \rho
k \to k \left[ 1+ \rho e^{-\alpha(h_n+h_{n-1})}\right],%(h_n-h_{n-1})^2
\label{stackPB}
\end{equation}
whose origin lies in the change in the electronic distribution on the bases
when the hydrogen bonds are broken and that provides a more realistic treatment of the phosphate backbone stiffness.
This new term leads to very sharp melting transitions at substantially reduced denaturation temperatures~\cite{PBD}.

Let us now briefly review the different scenarios that have been reported 
for the denaturation transition depending on the stiffness parameter $\rho$, for both homogeneous and heterogeneous DNA samples.
There is a consensus that, in the simplest case $\rho=0$, second-order denaturation transitions
are observed irrespective of the composition of the sample (whether heterogeneous or homogeneous)~\cite{cule,Theo}.
A nonvanishing $\rho$ and heterogeneous sequences successfully exhibit 
the characteristic abrupt, multistep melting observed in heterogeneous DNA molecules~\cite{cule} for intermediate-length sequences (see also~\cite{joyeux08}).
On the contrary, in the case of nonzero $\rho$ and homogeneous DNA the situation is much less clear. On the one hand,
there is a claim by Cule et al. that the transition region is extremely narrow, making it 
very sharp in appearance although, asymptotically, it is expected to be second-order~\cite{cule}.
On the other hand, Dauxois et al.~\cite{PBD} reported a first-order transition, yet with a diverging correlation length~\cite{Theo}.
Subsequent improved transfer-integral investigations by Joyeux et al.~\cite{joyeux05} did not settle the question, inasmuch as 
the numerics seems to indicate a first-order transition, but without discarding the possibility of a narrow second-order one.
Finally, results from a very recent path integral investigation suggest that the denaturation of homogeneous DNA has
the features of a second-order phase transition~\cite{zoli}.
In what follows, we clarify this situation by exploiting an analogy with the phenomenon of wetting.

\section{The wetting analogy}
\label{wetting}

The canonical partition function of the model factorises as usual into a product of kinetic and configurational parts,
$Z=Z_pZ_y$, with $Z_p=(2\pi m k_B T)^{N/2}$ and 
\begin{equation}
Z_y=\int  \prod_{n=0}^N dh_n e^{-\beta H'},
\end{equation}
where we have defined $H'=\sum_n( W+V)$ as the configurational part of $H$.
Given that most experiments on DNA thermal denaturation are performed in water,
the kinetic term does not play any role, and hence we can restrict ourselves on the 
configurational part.
In the continuum limit, for small values of $h_n -h_{n-1}$, 
the configurational part of the PBD Hamiltonian can be expressed as 
\begin{equation}
H_{ew}=\int dx \left[\frac{k}{2}(1+\rho e^{-2\alpha h})(\nabla h)^2 
+ w_1 e^{-ah} +w_2 e^{-2ah} \right]
\label{wetting_hamiltonian}
\end{equation}
where $w_1$, $w_2$ and $k$ are generic parameters. For $\rho=0$
$H_{ew}$ is the standard interfacial Hamiltonian for equilibrium critical wetting transitions 
in the presence of short-ranged forces, that is, the unbinding of the interface separating two 
coexisting phases from a substrate, which occurs upon increasing the temperature~\cite{review_wetting}. 
It constitutes an approximation to the PBD Hamiltonian that disregards the kinetic terms,
but from which equilibrium information can be gleaned.
The two strands of the DNA molecule correspond to the substrate and the interface in the wetting context,
and the denaturation of the former to the unbinding of the latter.
The analogy also extends to heterogeneous sequences, the thermal denaturation of heterogeneous DNA corresponding 
to the wetting of a one-dimensional interface from a disordered substrate.
This formal relation between wetting and DNA thermal denaturation was already noticed 
by Fisher~\cite{fisher} and exploited by Cule and Hwa~\cite{cule} and by Ares et al.~\cite{ares}. The analogy was carried further in~\cite{nosotros}, 
where it was pointed out that the reported critical exponents characterising the
DNA denaturation transition in the homogeneous case, $\h \sim |\delta|^{\beta}$ and $\xi \sim |\delta|^{-\nu}$ (where
$\delta=(T-T_c)/T_c$ and $\xi$ is the correlation length)~\cite{Theo}, 
are those of two-dimensional critical wetting, $\beta=-1$ and $\nu=2$~\cite{review_wetting}.
It was also shown by numerical simulations that the average stretching $\h$ diverges as $t^{1/4}$
at the transition temperature, in agreement with the exact result for the thickness
of the wetting layer~\cite{lipowsky}. Furthermore, 
the density of closed base-pairs~\cite{cule} scales as
the surface order-parameter in wetting, $\langle h^{-1} \rangle \sim |\delta|$~\cite{review_wetting}. 

Interestingly, it was also argued in~\cite{nosotros} that the theory of critical wetting should also apply to the 
PBD model (nonzero-$\rho$). Renormalisation group analyses of three-dimensional critical wetting
as embodied in eq. (\ref{wetting_hamiltonian}) with $\rho=0$ famously predict a strong non-universal critical behaviour~\cite{Brezin-FH}. These
predictions, however, are at odds with extensive Ising model computer simulations due to Binder et al.~\cite{binder} 
as well as with experiments~\cite{bonn}, which yield a mean-field-like second-order phase transition for the wetting problem.
Fisher and Jin~\cite{jf} suggested that this discrepancy arises from fundamental defects in the wetting Hamiltonian
\begin{equation}
H_{ew}(\rho=0)=\int dx \left[k (\nabla h)^2 + w_1 e^{-ah} +w_2 e^{-2ah} \right],
\label{hrho0}
\end{equation}
which should include a variable, position-dependent interfacial stiffness
\begin{equation}
k(h)=k+w'_1e^{-ah} +w'_2ah e^{-2ah}+ \cdots.
\end{equation}
When supplemented with the corrected stiffness, the structure of the wetting Hamiltonian eq.~(\ref{hrho0}) 
is very similar to that of the PBD one eq.~(\ref{wetting_hamiltonian})\footnote{
Just to complete the wetting story, the Fisher-Jin improved Hamiltonian did not yield the desired result, 
namely a crossover to mean-field like behaviour.
According to a linear renormalisation-group study, the presence of the
term proportional to $w'_2>0$ is capable of destabilising the critical
wetting transition, driving the transition weakly first-order
depending on system parameters in $d=3$~\cite{jf}. A subsequent
investigation allowed the analysis to be extended concluding that a
first-order transition can appear only for dimensions $d\geq 2.41$~\cite{boulter}. 
This puzzling situation was clarified only a few years ago by Parry et al.~\cite{parry}, who argued that 
the effective interfacial Hamiltonian for short-range critical wetting in three dimensions is in fact nonlocal, and that in the small
gradient limit, $\nabla h \ll 1$, it reduces to that proposed by Fisher et al.~\cite{jf}.
However, and remarkably, a thorough renormalisation-group  
and computer simulation analysis of the nonlocal Hamiltonian shows no stiffness instability
and hence the wetting transition remains continuous~\cite{parry}.}. A more detailed comparison 
reveals that in critical wetting the parameter $w'_1$ vanishes linearly with the transition temperature and
it is the next-to-leading term, $w'_2 e^{-2ah}$, that controls the critical behaviour. On the contrary,
for the PBD Hamiltonian $w'_1=k \rho /2>0$ is finite and $w'_2$ is identically zero. 
Thus, a renormalisation group analysis of the PBD model along the same lines as in the wetting case
requires switching on a nonvanishing $w'_1$ and truncating the series to first order. 
Despite these differences, we shall show that one can avail from 
standard renormalisation group techniques developed for the wetting problem to draw conclusions on the
order of the DNA denaturation transition. Such an analysis is carried out in the next section, where we prove 
that the one-dimensional melting transition for homogeneous DNA sequences is continuous in $\h$.

\section{Exact decimation procedure}
\label{rg}

The wetting analogy allows us to perform an exact decimation 
renormalisation-group (RG)~\cite{Nelson} analysis of the DNA denaturation 
in the PBD model. This RG procedure was successfully applied to the 2D wetting
transition~\cite{Julicher} ($H_{ew}(\rho=0)$) where the non-trivial fixed points can be 
analytically calculated for short-ranged forces~\cite{Huse}, and also for 
more general interactions~\cite{Spohn}. These calculations can be straightforwardly extrapolated
to the PBD model with $\rho\not=0$ which, as shown before, is formally equivalent to the discrete
version of the standard interfacial Hamiltonian for the 2D wetting transition
because the position-dependent stacking interaction in the PBD model is the analogous of 
the position-dependent stiffness in the wetting case.
Recent studies 
%of a quasi-one-dimensional model for the filling transition 
%on a linear wedge 
show that the dependence on the position
of the effective stiffness can induce new critical phenomena, and it can
even drive the transition first-order~\cite{Romero}.

The configurational part of the PBD Hamiltonian can be written as $H'=
\sum_{i=1}^N H_i^{(0)}(h_i,h_{i+1})$, where $H_i^{(0)}=W(h_i,h_{i+1})+(V(h_i)+
V(h_{i+1}))/2$. By simplicity we shall consider periodic boundary conditions 
$h_1=h_{N+1}$, although this choice will not affect our conclusions.
We define the initial transfer matrix as $T^{(0)}(h_i,h_{i+1})=
\exp(-\beta H_i^{(0)})$. 
Note that $T^{(0)}$ is symmetrical under an exchange of 
its arguments. The decimation RG procedure defines the transfer matrix 
$T^{(n)}$ at any RG step $n\ge 1$ as
\begin{eqnarray}
T^{(n)}(h,h')&=&b^\zeta \int_{-\infty}^{\infty}dh_1\ldots\int_{-\infty}^\infty
dh_{b-1}T^{(n-1)}(b^\zeta h,h_1)\nonumber\\
&\times&T^{(n-1)}(h_1,h_2)\ldots T^{(n-1)}(h_{b-1},b^\zeta h'),
\label{decimation}
\end{eqnarray}
where $b\ge 2$ is the rescaling factor and $\zeta$ is the wandering exponent. 
The normalised interaction is defined as 
$\beta H^{(n)}=-\ln T^{(n)}$. Note that this functional renormalisation 
preserves the symmetry under exchange of the arguments of $H^{(n)}$. 
This RG scheme is formally exact, but it cannot be solved analytically 
in general.
Instead of solving numerically the RG recursive equations, we shall analyse the
effect of the RG iterations in a subspace of the functional space 
$\{H^{(n)}\}$. In particular, we shall consider the transfer matrix to be
equal to the propagator corresponding to the continuum limit of the
PBD Hamiltonian $H_{ew}\equiv H_{ew}[h;k,\rho,\alpha,a,w_1,w_2]$ given by eq.~(\ref{wetting_hamiltonian})
\begin{equation}
T(h_0,h_1)\equiv Z(h_0,h_1;x=1)=\int {\mathcal D}h \textrm{e}^{-\beta H_{ew}},
\label{propagator}
\end{equation} 
where we integrate over all the continuum paths $h(t)$ ($0\le t \le x$) 
subject to the conditions $h(0)=h_0$ and $h(x)=h_1$. Due to the properties of
the transfer integral, the application of the RG scheme eq.~(\ref{decimation}) 
to this class of Hamiltonians yields 
\begin{equation}
Z'(h,h';1)= b^\zeta Z(b^\zeta h,b^\zeta h'; b),
\label{renormalized-z}
\end{equation}
where $Z'(h,h';1)$ is the transfer matrix associated with
a new continuum PBD Hamiltonian $H'_{ew}\equiv H'_{ew}[h';k',\rho',\alpha',a',
w_1', w_2']$. The renormalized Hamiltonian parameters are related to the 
original ones via
\begin{eqnarray}
k'&=&kb^{2\zeta-1}\quad,\quad \rho'=\rho \quad ,\quad   
\alpha'=b^\zeta \alpha,\nonumber\\ 
a'&=&b^\zeta a\quad,\quad
w_i'=w_i b, \quad i=1,2,...
\label{decimation2}
\end{eqnarray}
The wandering exponent is taken as $\zeta=1/2$ in analogy to the wetting case,
so $k$ and $\rho$ are unchanged by the RG iterations. On the other hand,
$a$, $\alpha$, $|w_1|$ and $w_2$ increase in each RG step. 
In order to reveal the irrelevance (in the RG sense) of the nonharmonic 
contribution of the stacking interaction, we shall show that the decimation 
procedure described above can be related to an analogous RG scheme for a 
modified PB model (i.e. with $\rho=0$). First, we note that the presence of 
the position-dependent term in the stacking interaction makes the definition
of the propagator eq.~(\ref{propagator}) ambiguous. A similar problem is
reported for the quantisation of classical systems with 
a position dependent mass~\cite{Thomsen,Chetouani}. We choose the following
definition of the propagator
\begin{equation}
Z(h_0,h_b;x)=\lim_{b\to \infty} \int dh_1\ldots dh_{b-1} \prod_{j=1}^b
K(h_j,h_{j-1};x/b),
\label{propagator2}
\end{equation} 
where $K(h,h';x)$ is defined as
\begin{eqnarray}
K(h,h';x)=\sqrt{\frac{\beta k(1+\rho e^{-\alpha(h+h')})}{2\pi x}}
\nonumber\\
\times \exp \left(-\frac{\beta k}{2x}(1+\rho e^{-\alpha(h+h')})(h-h')^2-x
\tilde{v}(h,h')\right),
\label{propagator3}
\end{eqnarray}
with a modified potential $\tilde{v}(h,h')$ given by~\cite{cule,Theo}
\begin{eqnarray}
\tilde{v}(h,h')=\beta w_1 e^{-\frac{a}{2}(h+h')} + \beta w_2 e^{-a(h+h')}
\nonumber\\
+ \frac{\beta w_3}{2}
\ln\left(1+\rho e^{-\alpha(h+h')}\right),
\label{modified-potential}
\end{eqnarray}
where initially $\beta w_3=1$.
With this definition the propagator can be understood as the result of a first
RG step for large $b$ and $x=b$ before rescaling the distance (see eq.~(\ref{decimation})). 
On the other hand, this expression reduces to the  
PB case as $\rho\to 0$. The propagator $Z$ verifies a Schr\"odinger-like 
equation. To obtain it, we note that for small $\Delta x$
\begin{eqnarray}
Z(h,h';x+\Delta x)\approx Z(h,h';x) + \Delta x \frac{\partial Z}{\partial x}
\nonumber\\
\approx \int dh'' Z(h,h'';x)K(h'',h';\Delta x).
\label{difeq1}
\end{eqnarray}
We expand $Z(h,h'';x)$ and $K(h'',h';\Delta x)$ for small $|h''-h'|$ as
\begin{widetext}
\begin{eqnarray}
Z(h,h'';x)\approx Z(h,h';x)+(h''-h')\frac{\partial Z(h,h';x)}{\partial h'}
+\frac{(h''-h')^2}{2} \frac{\partial^2 Z(h,h';x)}{\partial h'^2}, 
\label{difeq2}\\
K(h'',h';\Delta x)\approx G(h'';h',\Delta x)\Bigg[1+\Delta x \tilde v(h',h')
+ \frac{\Lambda^\prime(h')}{4\Lambda(h')}(h''-h')-
\frac{\Lambda^\prime(h')}{4\Delta x}(h''-h')^3 \nonumber \\
+ \frac{(h''-h')^2} {16} 
\left(\frac{\Lambda^{\prime\prime}(h')}{\Lambda(h')}-\frac{1}{2}
\left(\frac{\Lambda^\prime(h')}{\Lambda(h')}\right)^2\right)
%\nonumber\\
- \frac{(h''-h')^4}
{16\Delta x} \left(\Lambda^{\prime\prime}(h')+
\frac{(\Lambda^\prime(h'))^2}{\Lambda(h')}\right)
%\nonumber\\
+\frac{(h''-h')^6}{32(\Delta x)^2}(\Lambda^\prime(h'))^2
\Bigg],
\label{difeq3}
\end{eqnarray}
\end{widetext}
\begin{floatequation}
\mbox{\textit{see eqs.~(\ref{difeq2}) and~(\ref{difeq3})}} 
\end{floatequation}
where we defined $\Lambda (h')=\beta k(1+\rho e^{-2\alpha h'})$, $G(h'';h',\Delta x)=\sqrt{\Lambda(h')/2\pi \Delta x}\exp(-\Lambda(h')(h''-h')^2/2\Delta x)$
and the prime denotes differentiation with respect to the indicated arguments. Due to the
Gaussian form of $G(h'';h',\Delta x)$, we can evaluate trivially the integrals
on $h''$. In the limit $\Delta x \to 0$, the resulting expression reduces to
\begin{equation}
-\frac{\partial Z(h,h';x)}{\partial x}=-\frac{1}{2}\frac{\partial}{\partial 
h'}\left(\frac{1}{\Lambda}\frac{\partial Z(h,h';x)}{\partial h'}\right)
+ v^* Z
\label{schrodinger1}
\end{equation}
with the initial condition $Z(h,h';0^+)=\delta(h-h')$, and where\footnote{
Note that this Schr\"odinger-like eq.~(\ref{effective_potential1}) is not the same as the one proposed in 
Ref.~\cite{Theo}, which is in fact associated with a non-Hermitian Hamiltonian 
operator. Our approach preserves the self-adjointness of the corresponding 
Hamiltonian operator, as expected from the symmetric character of 
$K(h,h';\Delta x)$.}  
\begin{equation}
v^*(h)=\tilde{v}(h,h)+\frac{\Lambda^{\prime\prime}(h)}{8\Lambda^2(h)}-
\frac{1}{2 \Lambda(h)}\left(\frac{\Lambda^\prime(h)}{\Lambda(h)}\right)^2.
\label{effective_potential1}
\end{equation}
Introducing the change of variables~\cite{Yu}
\begin{eqnarray}
\eta&=&\int dh \sqrt{\Lambda(h)},
\label{changevar1} \\
\tilde{Z}(\eta,\eta';x)&=&
\Lambda(h)^{-1/4} \Lambda(h')^{-1/4}Z(h,h';x),
\label{changevar2}
\end{eqnarray}
eq.~(\ref{schrodinger1}) yields
\begin{equation}
-\frac{\partial \tilde{Z}(\eta,\eta';x)}{\partial x}=
-\frac{1}{2}\frac{\partial^2 \tilde{Z}(\eta,\eta';x)}{\partial \eta'^2}
+ v \tilde{Z}
\label{schrodinger2}
\end{equation}
with the effective potential $v=v(\eta)$ being defined as
\begin{eqnarray}
v(\eta)&=&v^*(h(\eta))+\left[\frac{7}{32 \Lambda}
\left(\frac{\Lambda^\prime}{\Lambda}\right)^2-\frac{\Lambda^{\prime\prime}}
{8\Lambda^2}\right]_{h=h(\eta)}
\nonumber\\
&=&\tilde v(h(\eta),h(\eta))-\frac{9}{8
\Lambda}\left(\frac{\Lambda^\prime}{\Lambda}\right)^2\Bigg |_{h=h(\eta)}.
\label{effective_potential2}
\end{eqnarray}
Consequently, the propagator of the PBD model can be mapped onto a propagator
of a PB-like model (i.e. $\rho=0$), where the effect of the position-dependent 
stacking interaction is absorbed into the definition of the variable $\eta$ 
and the effective potential $v$. From eq.~(\ref{changevar1}) we obtain the 
expression for the variable $\eta$
\begin{eqnarray}
\eta(h;\alpha,k,\rho)=
\sqrt{\beta k}\Bigg[h+\frac{1}{\alpha}\Big(\ln\frac{1+\sqrt{1+\rho 
e^{-2\alpha h}}}{2}\nonumber\\
-\sqrt{1+\rho e^{-2\alpha h}}+1\Big)\Bigg],
\label{expression_eta}
\end{eqnarray} 
so $\eta\sim \sqrt{\beta k} h$ for $h> \alpha^{-1}$. 
On the other hand, this expression verifies $\eta(\sqrt{b}h;\alpha,k,
\rho)=\sqrt{b}\eta(h;\sqrt{b}\alpha,k,\rho)$. Finally, the second 
term in eq.~(\ref{effective_potential2}) decays exponentially at large 
distances as 
\begin{equation}
\frac{9}{8\Lambda} \left(\frac{\Lambda^\prime}{\Lambda}\right)^2=
\frac{9\alpha^2\rho^2 e^{-4\alpha h}}
{8\beta k \left(1+\rho e^{-2\alpha h}\right)^3} 
\sim \frac{9\alpha^2\rho^2e^{-4\alpha h}}
{8\beta k}.
\label{effective_potential3}
\end{equation}
Therefore, the effective potential $v(\eta)$ decays exponentially with
$\eta$. Substituting eq.~(\ref{changevar2}) into eq.~(\ref{renormalized-z}), and
taking into account eq.~(\ref{expression_eta}), we find that under a RG
step the propagator $\tilde Z$ renormalizes as
\begin{eqnarray}
\tilde{Z}'&\equiv& \tilde{Z}'(\eta(h;\alpha',k',\rho'),\eta(h';\alpha,k,\rho);1)
\nonumber\\
&=&\sqrt{b}\tilde{Z}
(\eta(\sqrt{b}h;\alpha,k,\rho),\eta(\sqrt{b}h';\alpha,k,\rho);b)
\label{renormalized-z2}
\end{eqnarray}
where the renormalized effective potential parameters follow eqs.~(\ref{decimation2}) with $\zeta=1/2$. 
Note that eq.~(\ref{renormalized-z2})
is not the recursion relationship for 2D RG decimation scheme. 
However, as we iterate the RG equations, the variable $\eta$ becomes 
proportional to $h$ as $\alpha^{-1}\to 0$. Consequently, the high-temperature 
(HTFP) and critical (CFP) fixed points of the standard decimation RG procedure
$T^*_{\rm HTFP}$ and $T^*_{\rm CFP}$, respectively,~\cite{Huse,Burkhardt}
\begin{eqnarray}
T^*_{\rm HTFP}&=&\sqrt{\frac{\beta k}{2\pi}}
\left(e^{-\frac{\beta k(h-h')^2}{2}}-e^{-\frac{\beta k(h+h')^2}{2}}\right),
\label{fixed-points1}\\
T^*_{\rm CFP}&=&\sqrt{\frac{\beta k}{2\pi}}
\left(e^{-\frac{\beta k(h-h')^2}{2}}+e^{-\frac{\beta k(h+h')^2}{2}}\right),
\label{fixed-points2}
\end{eqnarray}
are also fixed points for the recursion eq.~(\ref{renormalized-z2}), 
corresponding to $\alpha^{-1}=0$. Moreover, the RG flow given by 
eq.~(\ref{renormalized-z2}) differs from the expression
for a true decimation only in terms proportional to $\exp(-2\alpha h)$ 
for large $\alpha$, which lead to irrelevant corrections in the RG sense. 
Therefore, the RG flow close to the critical fixed point must
be qualitatively similar to that obtained for $\rho=0$,  
and we conclude that the DNA denaturation transition in the PBD 
model is continuous and belongs to the 2D short-ranged, critical wetting 
universality class. 
 
\section{Conclusions}
\label{conclusions}
In this paper we have addressed the question of the order of the DNA 
denaturation transition for the PBD model. By using an exact decimation
procedure, we have shown that the position-dependent stacking interaction is
irrelevant in the RG sense, so the transition is continuous and in the
same universality class as the 2D critical wetting for short-ranged forces. 
However, our analysis only identifies the true asymptotic critical behaviour, not its range. 
% Hence, if the universal critical region is narrow enough, as it happens
% when $\alpha\ll a$~\cite{cule}, numerically obtained denaturation
% transitions in the PBD model may look like first-order, and this would
% explain the difficulties in determining the order of the transition
% that have been reported in the literature.

If the universal critical region is narrow enough, numerically obtained denaturation
transitions in the PBD model may look like first-order. For typical values of the parameters, 
a crossover temperature $T_{\rm cross}/T_m\sim 0.99$ 
has been numerically estimated, $T_m$ being the melting temperature (see Fig. 2 of Ref. \cite{cule}). 
Thus, the critical region turns out to be very narrow, which is a consequence of the entropic 
barrier induced by the anharmonicity in the stacking interaction \cite{cule}.
This would  explain the difficulties in determining the order of the transition
that have been reported in the literature.

\acknowledgments
J.M.R.-E. acknowledges financial support from Spanish Ministerio de Ciencia e 
Innovaci\'on and Junta de Andaluc\'ia through Grants FIS2009-09326 and 
P06-FQM-01869, respectively. A ``Ram\'on y Cajal'' Fellowship from the 
Spanish Ministerio de Ciencia e Innovaci\'on is also gratefully acknowledged.
F.dlS. and M.A.M. acknowledge financial support from Junta de Andaluc\'{\i}a 
through Grant FQM-165 and from the Spanish Ministerio de Ciencia e Innovaci\'on
through Grant FIS2009-08451.

\end{document}